\documentclass[conference]{IEEEtran}
\usepackage{packages}

\IEEEoverridecommandlockouts

\IEEEpubid{\makebox[\textwidth]{ISBN 978-3-903176-57-7\copyright{} 2023 IFIP \hfill}}%

\DeclareAcronym{cpu}{
    short               = {CPU},
    long                = {CPU}, %
    first-style         = short,
}

\DeclareAcronym{ack}{
    short               = {ACK},
    long                = {acknowledgment},
}

\DeclareAcronym{aes}{
    short               = {AES},
    long                = {Advanced Encryption Standard},
}

\DeclareAcronym{aesni}{
    short               = {AES-NI},
    long                = {Advanced Encryption Standard},
    first-style         = short,
}

\DeclareAcronym{chacha20}{
    short               = {ChaCha20},
    long                = {ChaCha20},
    first-style         = short,
}

\DeclareAcronym{qir}{
	short				= {QIR},
	long				= {QUIC Interop Runner},
}

\DeclareAcronym{io}{
    short               = {I/O},
    long                = {I/O},%
    first-style         = short,
}

\DeclareAcronym{udp}{
	short				= {UDP},
	long				= {UDP}, %
	first-style			= short,
}

\DeclareAcronym{tcp}{
    short               = {TCP},
    long                = {TCP}, %
    first-style         = short,
}

\DeclareAcronym{dpdk}{
    short               = {DPDK},
    long                = {Data Plane Development Kit},
}

\DeclareAcronym{tls}{
    short               = {TLS},
    long                = {Transport Layer Security},
    first-style         = short,
}

\DeclareAcronym{aead}{
    short               = {AEAD},
    long                = {authenticated encryption with additional data},
}

\DeclareAcronym{ietf}{
    short               = {IETF},
    long                = {Internet Engineering Task Force},
}

\DeclareAcronym{http}{
    short               = {HTTP},
    long                = {Hypertext Transfer Protocol},
    first-style         = short,
}

\DeclareAcronym{https}{
    short               = {HTTPS},
    long                = {Hypertext Transfer Protocol Secure},
    first-style         = short,
}

\DeclareAcronym{nic}{
    short               = {NIC},
    long                = {Network Interface Card},
}

\DeclareAcronym{dma}{
    short               = {DMA},
    long                = {Direct Memory Access},
}

\DeclareAcronym{tso}{
    short               = {TSO},
    long                = {TCP Segmentation Offload},
}

\DeclareAcronym{lro}{
    short               = {LRO},
    long                = {Large Receive Offload},
}

\DeclareAcronym{gso}{
    short               = {GSO},
    long                = {Generic Segmentation Offload},
}

\DeclareAcronym{gro}{
    short               = {GRO},
    long                = {Generic Receive Offload},
}

\DeclareAcronym{mss}{
    short               = {MSS},
    long                = {Maximum Segment Size},
}

\DeclareAcronym{nat}{
    short               = {NAT},
    long                = {Network Address Translation},
}

\DeclareAcronym{sctp}{
    short               = {SCTP},
    long                = {Stream Control Transmission Protocol},
}

\DeclareAcronym{rcvbuf}{
    short               = {RCVBUF},
    long                = {Receive Buffer},
}

\DeclareAcronym{sndbuf}{
    short               = {SNDBUF},
    long                = {Send Buffer},
}

\DeclareAcronym{cc}{
    short               = {CC},
    long                = {Congestion Control},
}

\DeclareAcronym{os}{
    short               = {OS},
    long                = {Operating System},
}

\begin{document}

\title{QUIC on the Highway:\\Evaluating Performance on High-rate Links}

\author{\IEEEauthorblockN{Benedikt Jaeger, Johannes Zirngibl, Marcel Kempf, Kevin Ploch, Georg Carle}
\IEEEauthorblockA{\textit{Technical University of Munich, Munich, Germany}\\
\{jaeger, zirngibl, kempfm, plochk, carle\}@net.in.tum.de\\
}
}

\maketitle

\begin{abstract}
    \quic is a new protocol standardized in 2021 designed to improve on the widely used \acs{tcp}\,/\,\acs{tls} stack.
The main goal is to speed up web traffic via HTTP, but it is also used in other areas like tunneling.
Based on \acs{udp} it offers features like reliable in-order delivery, flow and congestion control, stream-based multiplexing, and always-on encryption using \acs{tls}~1.3.
Other than with \acs{tcp}, \quic implements all these features in user space, only requiring kernel interaction for \acs{udp}.
While running in user space provides more flexibility, it profits less from efficiency and optimization within the kernel.
Multiple implementations exist, differing in programming language, architecture, and design choices.

This paper presents an extension to the \acl{qir}, a framework for testing interoperability of \quic implementations.
Our contribution enables reproducible \quic benchmarks on dedicated hardware.
We provide baseline results on 10G links, including multiple implementations, evaluate how \acs{os} features like buffer sizes and \acs{nic} offloading impact \quic performance, and show which data rates can be achieved with \quic compared to \acs{tcp}.
Our results show that \quic performance varies widely between client and server implementations from \SI{90}{\mega\bit\per\second} to \SI{4900}{\mega\bit\per\second}.
We show that the \ac{os} generally sets the default buffer size too small, which should be increased by at least an order of magnitude based on our findings.
Furthermore, \quic benefits less from \acs{nic} offloading and \acs{aes} NI hardware acceleration while both features improve the goodput of \ac{tcp} to around \SI{8000}{\mega\bit\per\second}.
Our framework can be applied to evaluate the effects of future improvements to the protocol or the \acs{os}.

\end{abstract}

\acresetall


\section{Introduction}
\label{sec:introduction}

\quic is a general-purpose protocol that combines transport layer functionality, encryption through \ac{tls} 1.3, and features from the application layer.
Proposed and initially deployed by Google~\cite{quicAnnouncement2013,langley2017quic}, it was finally standardized by the \ac{ietf} in 2021~\cite{rfc9000} after more than five years of discussion.
Like \ac{tcp}, it is connection-oriented, reliable, and provides flow and congestion control in its initial design.
An extension for unreliable data transmission was added in RFC9221~\cite{rfc9221}.

One goal of \quic is to improve web communication with \ac{https}, which is currently using \tcptls as underlying protocols.
It achieves this by accelerating connection build-up with faster handshakes, allowing only ciphers considered secure, and fixing the head-of-line blocking problem with \ac{http}/2.
The transport layer handshake is directly combined with a \ac{tls} handshake allowing 0- and 1-RTT connection establishment.
To comply with currently deployed network devices and mechanisms, \quic relies on the established and widely supported transport protocol \ac{udp}.
The usage of \ac{udp} allows to implement \quic libraries in user space.
Thus, \quic can initially be deployed without requiring new infrastructure, and libraries can be easily implemented, updated, and shipped.

On the one hand, this resulted in a variety of implementations based on different programming languages and paradigms~\cite{quicImplementations2021}.
On the other hand, previous efforts to optimize the performance of existing protocols have to be applied to new libraries, kernel optimizations cannot be used to the same degree, and the negative impact of encryption on performance has to be considered.
Additionally, the heterogeneity among the \quic implementations requires a consistent measurement environment for a reproducible evaluation.
The performance differences between \quic and \tcptls become more evident when the implementations are pushed to their limits on high-speed networks.

In this work, we show the impact of these effects through a fine-grained analysis of different \quic implementations.
We extend the existing \ac{qir}, a framework for interoperability testing of \quic implementations~\cite{seemann2020automating,seemann2020interop}.
Using Docker containers, it primarily focuses on functional correctness.
Thus, we extend it to allow for performance-oriented measurements on bare metal.

\vspace{0.2em}
\noindent
Our contributions in this paper are:

\first We develop and publish a \textbf{measurement framework based on the existing \ac{qir} framework}.
The measurements are run on real hardware, and more metrics are collected allowing an in-depth analysis of performance bottlenecks.
Besides our code, all configurations and the collected data are published along with this paper.
With this, we want to foster reproducibility and allow other researchers and library maintainers to evaluate different \quic libraries and test potential performance optimizations.

\second We perform a \textbf{baseline performance evaluation of different \quic implementations} on a 10G link with the proposed framework.
Tested implementations show a wide diversity in their configuration and behavior.

\third We \textbf{evaluate impacting factors on the performance} of \quic-based data transmissions.
We analyze the effect of, \eg different buffer sizes, cryptography, and offloading technologies.
This shows how goodput can be increased beyond the default setting.

\vspace{0.2em}
We explain relevant background regarding \quic in \Cref{sec:background}.
In \Cref{sec:framework}, we introduce the implemented measurement framework and used configurations.
We present our findings and evaluations in \Cref{sec:evaluation}.
\Cref{sec:related-work} contains an outline of related work.
Finally, we discuss our findings and conclude in \Cref{sec:conclusion}.

\section{Background}
\label{sec:background}
This section introduces relevant background for various properties impacting \quic performance, as shown in \Cref{sec:evaluation}.
We identify components relevant to the overall performance and point out the main differences compared to \tcptls.
They include the always-on encryption in \quic, the different \ac{ack} handling, the involved buffers in the network stack, and offloading functionalities supported by the \ac{nic}.

\subsection{Encryption}
\label{sec:background-encryption}
\quic relies on \ac{tls} version 1.3, which reduces available cipher suites to only four compared to previous \ac{tls} versions.
Only ciphers supporting \ac{aead} are allowed by the RFC~\cite{rfc9001}.
\ac{aead} encrypts the \quic packet payload while authenticating both the payload and the unencrypted header.
Supported ciphers either rely on \ac{aes}, a block cipher with hardware acceleration on most modern CPUs, or \ac{chacha20}, a stream cipher that is more efficient than \ac{aes} without hardware acceleration\footnote{\url{https://datatracker.ietf.org/doc/html/rfc8439\#appendix-B}}.

Considering \ac{tls} in combination with \ac{tcp}, data is encrypted into so-called \ac{tls} records, which are, in general, larger than individual \ac{tcp} segments spanning over multiple packets.
\ac{tcp} handles packetization and the reliable, in-order transfer of the record before it is reassembled and decrypted at the receiver.
With \quic, each packet must be sent and encrypted individually.
On receiving a packet, it is first decrypted before data streams can be reordered.
Loss detection and retransmissions are done on packet- and not stream-level using the packet number similar to \ac{tcp}'s sequence number.

Additionally, \quic adds another layer of protection to the header data, called header protection.
Fields of the \quic header like the packet number are encrypted again after packet protection has been applied, leading to twice as many encryption and decryption operations per packet.

\subsection{Acknowledgments}
Since \quic provides reliability and stream orientation, it requires a similar \ac{ack} process as \ac{tcp}.
\quic encapsulates \quic packets into \ac{udp} datagrams, while the packets carry the actual payload as \quic frames.
Different frame types exist, such as \textit{stream}, \textit{\ac{ack}}, \textit{padding}, or \textit{crypto}.

An \textit{\ac{ack}} frame contains so-called \ac{ack} ranges, acknowledging multiple packets and ranges simultaneously, potentially covering multiple missing packets.
All received packets are acknowledged.
However, \textit{\ac{ack}} frames are only sent after receiving an \ac{ack}-eliciting packet.
A \quic packet is called \ac{ack}-eliciting if it contains at least one frame other than \textit{\ac{ack}}, \textit{padding}, or \textit{connection close}.

Other than \ac{tcp}, \quic sends \acp{ack} encrypted, inducing additional cryptographic workload both on sending and receiving side.
Therefore, \quic must determine how frequently \acp{ack} are sent.
The \textit{maximum \ac{ack} delay} transport parameter defines the time the receiver can wait before sending an \textit{\ac{ack}} frame~\cite{rfc9000}.
Determining the acknowledgment frequency is a trade-off and may affect the protocol's performance.
Fewer \acp{ack} can lead to blocked connections due to retransmissions and flow control, while more put additional load on the endpoints.

Furthermore, sending and receiving \acp{ack} with \quic is more expensive than with \ac{tcp}.
With \ac{tcp}, \acp{ack} do not take any additional space since they are part of the \ac{tcp} header and thus can be piggybacked on sent data.
The kernel evaluates them before any cryptography is performed.
We show the impact of \acl{ack} processing and sending in \Cref{sec:evaluation}.

\subsection{Buffers}
\label{sec:background-buffer}
\begin{figure}
	\centering
	\includegraphics{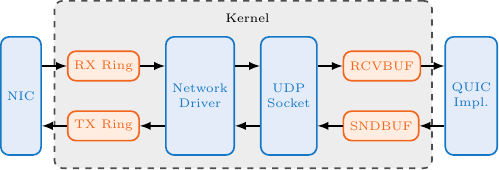}
	\caption{Simplified overview of available system buffers relevant for a \quic implementation.}
	\label{fig:buffers}
\end{figure}

Different buffers of the system can impact the performance of \quic.
\Cref{fig:buffers} shows a simplified schema of buffers managed by the kernel, which are relevant for the measurement scenarios of this paper.
The \ac{nic} stores received frames to the RX ring buffer using \ac{dma} without kernel interrupts.
Conversely, it reads frames from the TX buffer and sends them out.
The kernel reads and writes to those ring buffers based on interrupts and parses or adds headers of layers 2 to 4, respectively.
Afterward, the UDP socket writes the payload to the \ac{rcvbuf}.

However, if the ring buffer or \ac{rcvbuf} is full, packets are dropped, and loss occurs from the perspective of the \quic implementation.
The default size of the receive buffer in the used Linux kernels is \SI{208}{\kibi\byte}.
In contrast, if the \ac{sndbuf}, which resides between the implementation and the socket, is packed, the \quic implementation will be blocked until space is available.
Both scenarios reduce the goodput.
We show the impact of different buffer sizes in \Cref{sec:buffers}.

\subsection{Offloading}
The Linux kernel offers different \ac{nic} offloading techniques, namely \ac{tso}, \ac{gso}, and \ac{gro}.
While the first only affects \ac{tcp}, the others also apply to \ac{udp} and thus \quic.
The main idea of all offloading techniques is to combine multiple segments and thus reduce the overhead of processing packets.
Following the ISO\,/\,OSI model, data sent by \ac{tcp} is supposed to be split into chunks smaller or equal to the \ac{mss} and then passed down to lower layers.
Each layer adds its header and passes the data further.
All of this is computed by the kernel.
With hardware offloading, this segmentation task can be outsourced to the \ac{nic}.
Offloading can only accelerate sending and receiving of packets, not the data transfer over the network~\cite{kerneloffload}.

So far, most offloading functions are optimized regarding \ac{tcp}.
\quic profits less since packetization is done in user space.
Utilizing offloading would require adjustments in the implementation to offload tasks like cryptography or segmentation~\cite{yang2020quicnicoffloading}.
In \Cref{sec:offloading} we compare the impact of different offloading functions on \quic and \tcptls.

\section{Measurement Framework}
\label{sec:framework}

\begin{figure*}
	\centering
	\includegraphics[]{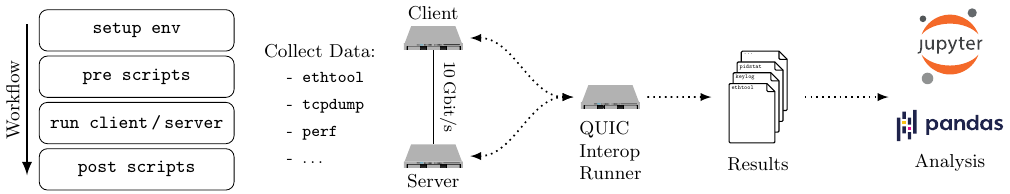}
	\caption{Hardware architecture, measurement workflow, and analysis pipeline.}
	\label{fig:architecture}
\end{figure*}

In this paper, we extend \ac{qir} initially proposed by \citet{seemann2020automating}.
It is designed to test different \quic implementations for interoperability and compliance with the \quic standard reporting results on a website~\cite{seemann2020interop}.
Servers and clients from multiple implementations are tested against each other, performing various test cases, like handshake, 0-RTT, and session resumption.
The main focus is on the implementations' functional correctness and less on performance.
Even though all followed the same RFC drafts during specification, interoperability between different servers and clients was only present for some features.
Implementations run inside Docker containers, and the network in between is simulated using \mbox{\textit{ns-3}}\footnote{\url{https://www.nsnam.org/}}.
Simple goodput measurements are available but only conducted on a \SI{10}{\Mbps} link.
As of January 2023, nearly all tested libraries are close to the possible maximum goodput~\cite{seemann2020interop}.

\ac{qir} orchestrates the measurements by configuring used client and server applications, creating required directories for certificates and files, and collecting log files and results afterward.
In the sense of reproducibility, these features make \ac{qir} a powerful tool for \quic (but not limited to) evaluations.

We extended \acs{qir} to enable high-speed network measurements on dedicated hardware servers.
\Cref{fig:architecture} shows an overview of our framework and its features.
For this work, we use different physical servers for the client and server implementation to prevent them from impacting each other.
We add additional configuration parameters, such that measurements can be specified with a single configuration file, and include version fingerprints both from the implementations and \ac{qir} to the output to foster reproducibility.
Additionally, we extend the logging by including several tools which collect data from different components (\eg the \ac{nic}, \ac{cpu}, and sockets).

We follow these three requirements in the implementation of our measurement framework:

\textbf{Flexibility:}
Any \quic implementation can be used as long as it follows the required interface regarding how client\,/\,server are started and variables are passed to them. We provide example implementations for the ones given in \Cref{sec:implementations}.

\textbf{Portability:}
The framework can be deployed to any hardware infrastructure, as depicted in \Cref{fig:architecture}, requiring a link between the client and server nodes and \texttt{ssh} access from the management node.

\textbf{Reproducibility:}
Experiments are specified in configuration files and results contain needed information on how they were generated, \eg the \ac{qir} and \quic implementation version.
Additionally, results contain a complete description of the used configuration, versions, and hardware.

Our code, results, and analysis scripts are available:
\begin{center}
	\url{https://github.com/tumi8/quic-10g-paper}
\end{center}
This includes our extension of \ac{qir}, all measurement configurations, and results shown in the paper, analysis scripts to parse the results, and Jupyter notebooks for visualization~\cite{code-publication}.

\subsection{Workflow}
We followed a similar workflow as \ac{qir}.
First, our framework configures the used hardware nodes, especially the used network interfaces for the measurements.
Each measurement executes four different scripts that the implementation or configuration can adjust: setup environment, pre-scripts, run-scripts, post-scripts as shown in \Cref{fig:architecture}.
The setup script can be used to install local dependencies like Python environments.
Pre- and post-scripts can be utilized to configure OS-level properties like the \ac{udp} buffer size and reset them after the measurement.
Also, additional monitoring scripts can be started and stopped accordingly.

Relevant configuration parameters for the client\,/\,server implementation, such as IP address, port, and X.509 certificate files, are passed via environment variables.
Then the server application is started, waiting for incoming connections.
Subsequently, the client application conducts a \quic handshake and requests a file from the server via HTTP/3.
Once the client terminates, the framework checks whether all files were transferred successfully and executes post-scripts to stop monitoring tools and to reset the environment to a clean state for subsequent measurements.
Finally, the framework collects all logs.

\subsection{Hardware Configuration}
We decided to run the client and server applications on different hardware nodes to prevent interference and to fully include the kernel and \ac{nic} (other than with Docker).
The \ac{qir} runs on another host and functions as a management node orchestrating the measurement.
The management node can access the measurement nodes via \texttt{ssh}, while the others are connected via a \SI{10}{\Gbps} link, as shown in \Cref{fig:architecture}.
Different network interfaces and links are used for management and measurements.
If not stated differently, all measurements were executed on \textit{AMD EPYC 7543 32-Core} processors, \SI{512}{\giga\byte} memory, and \textit{Broadcom BCM57416 \acp{nic}}.
As an operating system, we use \textit{Debian Bullseye} on \textit{5.10.0-8-amd64} for all measurements without additional configurations.
We rely on live-boot images with RAM disks, drastically increasing \ac{io} speed to focus on the network aspect.

\subsection{Collected Data and Analysis}
For each measurement, the framework computes the goodput as the size of the transmitted file divided by the time it takes to transmit it.
To reduce this impact of transmission rate fluctuations (\eg caused by congestion control), a large file size (\SI{8}{\giga\byte}) is chosen to enforce a connection duration of several seconds.

The following monitoring tools are directly integrated into our framework to collect more data from within the implementations or from the hardware hosts.
\textbf{tcpdump} is used to collect packet traces which can be decrypted with the exported session keys.
Additionally, implementations can enable \textbf{qlog} (a schema for logging internal \quic events \cite{draft-ietf-quic-qlog-main-schema-04}) and save results to a directory set up and collected by the framework.
However, both result in extensive logging that heavily impacts performance and are only considered for debugging.
\textbf{ifstat}, \textbf{ethtool}, \textbf{netstat}, and \textbf{pidstat} can be started to collect additional metrics from the \ac{nic} and \ac{cpu}, such as the number of dropped packets or context switches.
Finally, \textbf{perf} can be used for in-detail client and server application profiling.
This allows to analyze how much \ac{cpu} time is consumed for tasks like encryption, sending, receiving, and parsing packets.
When enabled, the output of the used tools is exported along with the measurement results.

We provide a parsing script for all generated data that handles all different output formats of the used tools.
By this, it is possible to analyze the change of goodput with different configurations and get a better understanding of why this is the case.
It is possible to detect client and server-side bottlenecks as root causes for performance limitations.
We consider this important since we observed multiple different \quic components limiting the performance depending on the configuration.
The analysis pipeline parses available result files and outputs the final results as Python \textit{pandas} DataFrame.

\subsection{Implementations}
\label{sec:implementations}
Since there is not one widely used implementation for \quic (such as Linux for \ac{tcp}),
we evaluate multiple implementations written in different languages.
The implemented framework currently includes six \quic libraries namely:
\aioquic~\cite{aioquic},
\quicgo~\cite{quic-go},
\mvfst by Facebook~\cite{mvfst},
\picoquic by Private Octopus~\cite{picoquic},
\quiche by Cloudflare~\cite{quiche},
and \lsquic by LiteSpeed Tech~\cite{lsquic}.
For each implementation, we either used provided examples for the client and server or made minor adjustments to be compatible with \ac{qir}.

In this paper, we mainly focus on \lsquic and \quiche since they show the highest goodput rates, as shown in \Cref{sec:baseline}.
They are implemented in Rust and C, respectively, and are already widely deployed~\cite{zirngibl2021over9000}.
Furthermore, both libraries rely on BoringSSL for \ac{tls}.
Thus, cryptographic operations and TLS primitives are comparable, and we can focus on \quic specifics in the following.
We compare them to the remaining libraries in \Cref{sec:baseline} to put their initial performance into perspective and further support our selection.
We base our initial client and server on their example implementation and adapt only where necessary.
Additionally, we compare \quic with a \tcptls stack consisting of a server using \texttt{nginx} (version 1.18.0) and a client using \texttt{wget}.

\section{Evaluation}
\label{sec:evaluation}
We apply the measurement framework to evaluate the performance of different \quic implementations.
For every measurement, the client downloads an \SI{8}{\giga\byte} file via \ac{http}/3.
Every measurement is repeated 50 times to assure repeatability.
The boxplots show the median as a horizontal line, the mean as icons such as $\blacktriangle$, and the quartiles $Q_1$ and $Q_3$.
All implementations come with different available \ac{cc} algorithms.
For \lsquic, we observed unintended behavior with \textit{BBR}, resulting in several retransmissions and only between \SIrange{30}{70}{\percent} of the goodput of \textit{Cubic}.
We assume this is related to a known issue with \textit{BBR} in \ac{tcp}~\cite{scholz2018towards}.
To prevent significant impact from different algorithms, we set each implementation to \textit{Cubic} if available or \textit{Reno} otherwise.

\subsection{Baseline Measurements}
\label{sec:baseline}
\begin{figure}[tb]
    \centering
    \includegraphics{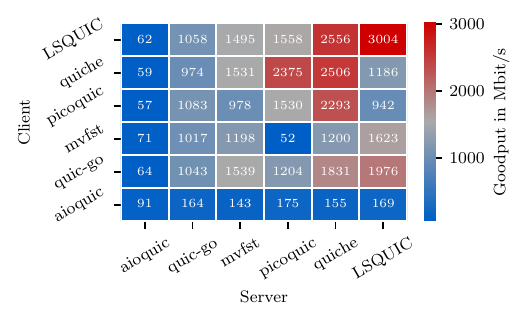}
    \caption{Baseline goodput results for different \quic libraries tested against each other on a \SI{10}{\giga\bit\per\second} link. In comparison \tcptls reaches \SI{8010}{\mega\bit\per\second} in the baseline scenario.}
    \label{fig:baseline}
\end{figure}

\ac{qir} is designed to test operability between different implementations.
As a first baseline, we evaluate the client and server of the implementations listed in \Cref{sec:implementations}.
\Cref{fig:baseline} shows the goodput for each client-server pair for all implementations.
It clearly shows that the goodput widely differs (\SIrange{52}{3004}{\Mbps}) between the implementations.

The Python library \aioquic shows the lowest performance, as expected since it is the only interpreted and not compiled implementation.
It can be considered an implementation suitable for functional evaluation or a research implementation.
When it comes to goodput, it can be neglected.
The two best-performing implementations are \lsquic and \quiche.
Regarding \picoquic, we are not able to reach similar goodput as reported by \citeauthor{tyunyayev2022picoquichighspeed} \cite{tyunyayev2022picoquichighspeed} and were not able to easily reproduce the required optimizations.
Therefore, we stick with \lsquic-\lsquic and \quiche-\quiche pairs for the following evaluation since they achieved the highest goodput.

Like interoperability of \quic features, goodput performance widely varies among client-server pairs.
We conjecture that different operation modes, \quic parameters, and efficiency of the used components result in these fluctuations.
The \lsquic server achieves the highest rate with the \lsquic client while being less performant with clients of other implementations.
Overall, the \quiche server achieves decent goodput with most clients.
The \picoquic-\mvfst measurements achieve peculiar low rates of only \SI{52}{\mega\bit\per\second}.

\noindent
\textit{\textbf{Key take-away:}
The goodput of different \quic libraries varies drastically, not only between libraries in general but also between clients and servers.
\lsquic and \quiche perform best in the baseline scenario.
Only in the case of \lsquic as server and \quiche as client the goodput drops drastically.
}

\subsection{Performance Profiling}
\label{sec:profiling}

\begin{figure}[tb]
	\centering
	\includegraphics{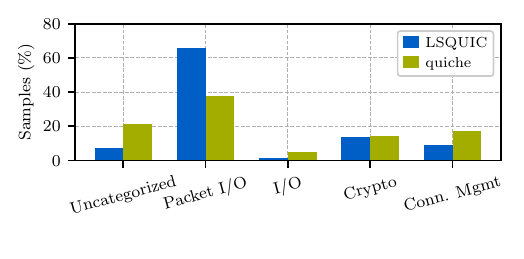}
	\caption{Distribution of server perf samples across different categories. The results are shown in relation to the total number of samples collected by perf (including idle states).}
	\label{fig:perf}
\end{figure}

We use \textbf{perf} to analyze \lsquic and \quiche further and see the \ac{cpu} time consumption for each component.
During each measurement, perf collects samples for the complete system.
We categorize the samples for both implementations based on their function names and a comparison to the source code.
\Cref{fig:perf} shows the results for the respective servers.
\textit{Packet \ac{io}} covers sending and receiving messages, especially the interplay with the \ac{udp} socket and kernel functionality (sendmsg and recvmsg).
\textit{\ac{io}} covers reading and writing the transmitted file by server and client.
\textit{Crypto} describes all \ac{tls}-related en-\,/\,decryption tasks.
\textit{Connection Management} covers packet and acknowledgment processing and managing connection states and streams.
If no function name can be collected by \textbf{perf} or if we cannot map it to a category, we use \textit{Uncategorized}.

We manually created this mapping by assigning each function occurring in the perf dump to a category.
While we could not assign all functions, a clear trend is visible.
The strict inclusion of \ac{tls} is often criticized due to the expectation of a high overhead~\cite{quic-crypto-discussion}.
However, \textit{Packet \ac{io}} takes up a majority of resources for both libraries.
Passing packets from the libraries in user space to the kernel, the \ac{nic}, and vice versa currently impacts the performance most.

Interestingly, we see differences in performance in the used programming language or architecture and between the two client and server combinations of different implementations.
Resulting in the highest difference, the goodput with a \quiche server and \lsquic client is more than twice as high as vice versa.
During these measurements, both the client and server are only at around \sperc{70} \ac{cpu} usage, no loss is visible, and no additional retransmission occurs.
In this case, the bottleneck seems to be the interaction of the flow control mechanisms of both libraries in this client\,/\,server constellation.
Besides general interoperability tests, our measurement framework can be used in the future to identify these scenarios and to improve \quic libraries, their flow and congestion control mechanisms, and their interaction.

\noindent
\textit{\textbf{Key take-away:}
    Our findings show that the most expensive task for \quic is Packet \ac{io}.
    While the cost of crypto operations is visible, it is not the main bottleneck here.
    Overall, our results comply with related work~\cite{yang2020quicnicoffloading}, and we share our mappings so that they can be refined and extended in the future.
}

\subsection{Buffers}
\label{sec:buffers}

As explained in \Cref{sec:background-buffer}, different system buffers affect a \quic connection.
The essential buffers are the ring buffers in-between the \ac{nic} and network driver, and the send and receive buffers from the socket used by the library, as shown in \Cref{fig:buffers}.

To analyze the impact of these buffers, the measurement framework captures network driver statistics before and after each data transmission using \textbf{ethtool} and connection statistics using \textbf{netstat}.
It provides the number of sent and received packets and dropped packets in each of the mentioned buffers.

\begin{figure}[tb]
    \centering
    \includegraphics[]{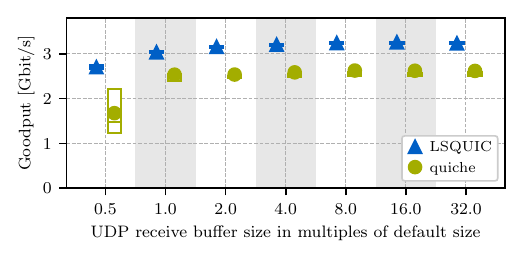}
    \caption{\quic goodput with different UDP receive buffer sizes. X-axis values are multiples of the default buffer size of \SI{208}{\kibi\byte}.}
    \label{fig:udp-buffer}
\end{figure}

The baseline measurement shows that the ring buffers drop no data.
However, packets are dropped by the client receive buffer since both client implementations retrieve packets slower than they arrive.
As a result, retransmissions are required by the server, and congestion control impacts the transmission.
To analyze the impact on the goodput, we increase the default receive buffer (\SI{208}{\kibi\byte}).
The effect of this can be seen in \Cref{fig:udp-buffer}.
It shows the goodput of \lsquic and \quiche pairs for different buffer sizes as multiple of the default buffer on the x-axis.
The goodput for both libraries improves with increasing buffer sizes and stabilizes after a 16-fold increase.

\begin{figure}[tb]
    \centering
    \includegraphics[]{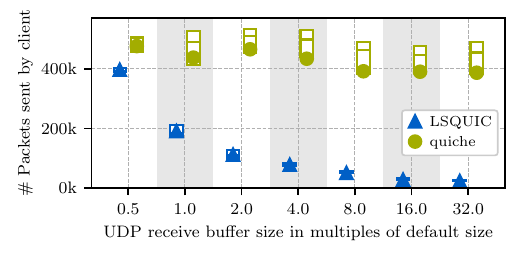}
    \caption{Number of sent ACKs by the client. X-axis values are multiples of the default buffer size of \SI{208}{\kibi\byte}.}
    \label{fig:udp-buffer-sent-acks}
\end{figure}

Using \lsquic with the default buffer size results in \sk{11.4} dropped packets and a loss rate of \sperc{0.2}.
With the largest tested buffer size, \lsquic reaches a goodput of \SI{3250}{\mega\bit\per\second}, an \sperc{8.7} increase compared to the baseline shown in \Cref{fig:baseline}.
Besides the reduced loss and retransmissions, the number of \acp{ack} sent by the \lsquic client is drastically reduced from \sk{180} to \sk{46} (\sperc{26}).
This development for all tested buffer sizes is shown in \Cref{fig:udp-buffer-sent-acks}.
The reduction of sent \ac{ack} packets also results in reduced \ac{cpu} usage by the client while increasing the server \ac{cpu} usage shifting the bottleneck further towards the sending of \quic packets.
In all scenarios, \lsquic sends fewer \acp{ack} than reported by \citet{marx2020implementationdiversity}.

Regarding \quiche, only \sk{7} packets are dropped with the default buffer size, hence a \sperc{0.1} loss rate.
Larger buffers decrease the loss by one order of magnitude and increase the goodput but only by \sperc{3} to \SI{2530}{\mega\bit\per\second}.
In contrast to \lsquic, no difference regarding sent \acp{ack} can be seen (see \Cref{fig:udp-buffer-sent-acks}).
Reducing the receive buffer further impacts both libraries, especially \quiche.
The goodput drops by \sperc{40} and a more significant deviation is visible.
We repeated the baseline measurements with larger \ac{udp} receive buffers for which the results can be seen in \Cref{fig:baseline_optimized}.

\noindent
\textit{\textbf{Key take-away:}
The default \ac{udp} receive buffer size has a visible impact on \quic-based data transmission.
We recommend a general increase of the default buffer by at least an order of magnitude to support widespread deployment of \quic.}

\subsection{Crypto}
\label{sec:crypto}
As explained in \Cref{sec:background-encryption}, \quic supports \ac{tls} using \ac{aes} or \ac{chacha20}.
Generally, operations to encrypt and decrypt are computationally expensive but are widely optimized in hardware and software today.
\quic and \ac{tls}~1.3 only support \ac{aead} algorithms~\cite{rfc8446}, which can encrypt and sign data in one single pass.
For \ac{aes}, only three ciphers are available:
\texttt{AEAD\_AES\_128\_GCM},
\texttt{AEAD\_AES\_128\_CCM}, and
\texttt{AEAD\_AES\_256\_GCM}.
From them, only \texttt{GCM} ciphers are required or should be implemented \cite{rfc8446}, the \texttt{CCM} cipher is rarely used.
Usually, the \SI{128}{\bit} \texttt{GCM} cipher is preferred~\cite{zirngibl2021over9000}.
For the following evaluation, we use the default cipher.
\ac{chacha20} always uses the \texttt{Poly1305} authenticator to compute message authentication codes.

We evaluated the \quic implementations in three different scenarios.
Two of them use \ac{aes} either without or with the \ac{aes} New Instruction Set that offers full hardware acceleration \cite{intelaesni} (which is the default in our test environment), and one scenario uses \ac{chacha20}.
For the following, we refer to the hardware-accelerated \ac{aes} version as \ac{aesni}.
\lsquic and \quiche automatically fall back to \ac{chacha20} whenever hardware acceleration is unavailable.
To evaluate \ac{aes} without hardware acceleration, we patched the used \texttt{BoringSSL} library accordingly.

\begin{figure}[tb]
    \centering
    \includegraphics[]{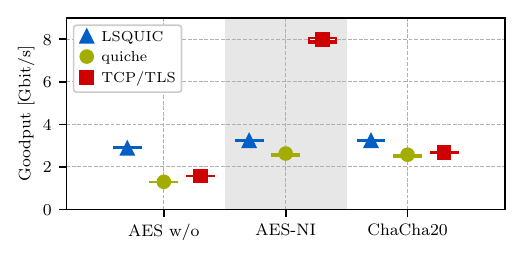}
    \caption{Impact of different \ac{tls} ciphers on \quic and \tcptls goodput. During the \ac{aes} measurement without hardware acceleration, the implementations were forced to use \ac{aes} by deactivating \ac{chacha20} in the respective \ac{tls} library to prevent the fallback.}
    \label{fig:crypto}
\end{figure}

As seen in \Cref{fig:crypto}, \ac{chacha20} achieves the same throughput as \ac{aesni} for both \quic libraries.
While removing hardware acceleration decreases the goodput with \ac{aes} by \SI{11}{\percent} for \lsquic and even by \SI{51}{\percent} for \quiche.
This difference between the implementations results from the \quiche client sending more than 16 times as many \acp{ack} than \lsquic.
While the server is the bottleneck in all measurements, more packets sent by the client add additional load to the server for receiving and decrypting packets (see \Cref{sec:buffers}).
This decryption is more expensive without hardware acceleration. 
Thus, crypto has a higher impact on the overall goodput.
Also, we observe that the client CPU utilization of \quiche is \SI{5}{\percent} lower with \ac{chacha20} than with \ac{aes}.

With \tcptls, the effect changes.
While \ac{chacha20} reaches higher goodput rates than \ac{aes} without hardware acceleration,
\ac{aesni} outperforms \ac{chacha20} with a three times higher goodput, almost reaching the link rate.
Additionally, for \tcptls we observed that the client CPU bottlenecked the connection for \ac{aesni} and \ac{chacha20}, while the server was the bottleneck without hardware acceleration.

\noindent
\textit{\textbf{Key take-away:}
    While selecting an appropriate cipher suite drastically impacts \tcptls, \quic reaches similar goodput for \ac{aesni} and the \ac{chacha20}-based cipher.
The main bottleneck is still packet \ac{io} and the acceleration effect is further reduced due to smaller \ac{tls} records and encrypted \acp{ack}.
}

\subsection{Segmentation Offloading}
\label{sec:offloading}
To reduce the impact of packet \ac{io}, a \quic library can combine multiple packets and rely on offloading.
Similarly, an implementation can use \textit{sendmmsg}\footnote{\url{https://man7.org/linux/man-pages/man2/sendmmsg.2.html}} and \textit{recvmmsg}\footnote{\url{https://man7.org/linux/man-pages/man2/recvmmsg.2.html}} to reduce system calls and move multiple packets to/from the kernel space within a single buffer.

\begin{figure}[tb]
    \centering
    \includegraphics[]{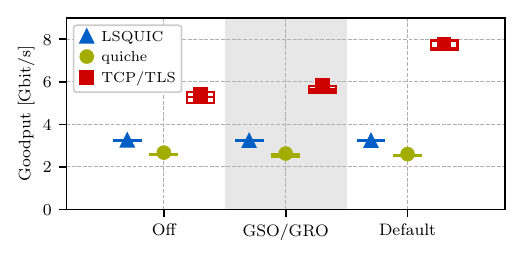}
    \caption{Impact of hardware offloading on \quic and \tcptls goodput. By default, \ac{gso}, \ac{gro}, and \ac{tso} are activated.}
    \label{fig:offloading}
\end{figure}

By default, \acf{gso}, \acf{gro}, and \acf{tso} are activated in Linux.
We analyze which offload impacts \quic and \ac{tcp} by incrementally activating different offloading techniques.
\Cref{fig:offloading} indicates that all the offloading techniques hardly influence \quic goodput, while \ac{tcp} largely profits from \ac{tso}.
The \quic goodput does not change when \ac{gso}\,/\,\ac{gro} is enabled.
However, the client CPU utilization increases from \SI{82}{\percent} to \SI{92}{\percent} for \lsquic and from \SI{77}{\percent} to \SI{84}{\percent} for \quiche.
Turning off all offloads does not decrease goodput but decreases the client's \ac{cpu} utilization and also power consumption.

While \lsquic implements functionality to support \textit{sendmmsg} and \textit{recvmmsg}, we were not able to use it as of January 2023.
Data transmission with the functionality activated randomly terminated with exceptions.
We expect improved support for those features also by other implementations and suggest reevaluating libraries with our framework in the future.

\noindent
\textit{\textbf{Key take-away:}
The tested \quic implementations do not profit from any segmentation offloading techniques as of January 2023.
Compared with \ac{tcp}, there is much room for improvement to apply the same benefits to \ac{udp} and \quic.
Since \quic encrypts every packet individually, including parts of the header, techniques similar to \ac{tso} cannot be applied, which would require that headers can be generated in the offloading function.
However, with adjustments to the protocol and the offloading functions, speedups can be achieved for segmentation and crypto offloading~\cite{yang2020quicnicoffloading}.
}

\subsection{Hardware}
\label{sec:evaluation-hardware}
Even though an implementation in user space comes with more flexibility, tasks that are done by the kernel for \ac{tcp} become more expensive with \quic.
More CPU cycles are required per packet.

We repeated the final goodput measurements with increased buffers on five host pairs with different generations of AMD and Intel \acp{cpu} listed in \Cref{tab:used-hardware}.
Client and server hosts are equipped with the same \ac{cpu} for each measurement.
Additionally, we optimized \lsquic and \quiche further by adding compile flags to optimize for the used architecture.
The optimized implementations are referred to as \lsquicopt and \quicheopt.

\begin{table}[tb]
	\centering
    \caption{Different CPUs used for the measurements. The default for measurements was AMD-2 if not noted otherwise.}
	\label{tab:used-hardware}
    \begin{tabular}{l l l r}
		\toprule
          & \ac{cpu} & Year & max \si{\giga\hertz}\\
		\midrule
        AMD-1 & AMD EPYC 7551P & 2017 & 2.0 \\
        AMD-2 & AMD EPYC 7543  & 2021 & 3.7 \\
        Intel-1 & Intel Xeon CPU D-1518  & 2015 & 2.2 \\
        Intel-2 & Intel Xeon CPU E5-1650 & 2012 & 3.8 \\
        Intel-3 & Intel Xeon Gold 6312U &  2021 & 3.6 \\
		\bottomrule
	\end{tabular}
\end{table}

\begin{figure}[t]
    \centering
    \includegraphics[]{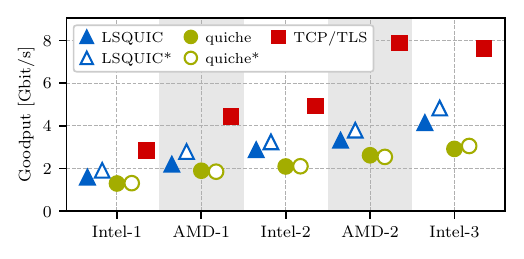}
    \caption{Goodput as measured on different hardware architectures listed in \Cref{tab:used-hardware}. \lsquicopt and \quicheopt are built with compile flags to optimize for the used architecture.}
    \label{fig:hardware}
\end{figure}

The results in \Cref{fig:hardware} show that quantizing \quic throughput highly depends on the used CPU and architecture.
Both \quic and \ac{tcp} profit from newer CPUs with more modern instruction sets.
Compile flags improved the \lsquic goodput by \SIrange{14}{20}{\percent}, while they hardly affected \quiche.
Also, \quic and \ac{tcp} perform differently among the different architectures.
For example, when comparing the two newest CPUs AMD-2 and Intel-3, \quic performs \SI{27}{\percent} better on the Intel chip, almost reaching \SI{5}{\Gbps}.
On the other side, \ac{tcp} goodput decreases by \SI{3}{\percent} compared to the \ac{cpu} in the AMD-2 host pair.
This shows that \quic (in the user space) and \ac{tcp} (in the kernel) profit differently from \ac{cpu} architectures and instruction sets.

\noindent
\textit{\textbf{Key take-away:}
The used hardware is highly relevant for evaluating \quic performance.
Newer \acp{cpu} lead to a higher goodput for both \quic implementations even though their frequency is slightly lower, \eg comparing Intel-2 and Intel-3.
While not feasible for all research groups, we suggest attempting an evaluation of potential improvements to \quic libraries on different \ac{cpu} generations and frequencies in the future to better quantify their impact.
}

\section{Related Work}
\label{sec:related-work}

The interoperability of different \quic implementations has been tested by research throughout the protocol specification.
\citet{seemann2020automating}, \citet{PirauxQuicEvolution2018}, and \citet{marx2020implementationdiversity} developed different test scenarios and analyzed a variety of QUIC implementations.
Furthermore, \citet{ruethQUIC2018}, \citet{PirauxQuicEvolution2018}, and \citet{zirngibl2021over9000} have already shown a wide adoption of \quic throughout the protocol specification and shortly before the release of RFC9000~\cite{rfc9000}.
They show that multiple large corporations are involved in the deployment of \quic, various implementations are used, and different configurations can be seen (\eg transport parameters).
However, they mainly focused on functionality analyses, interoperability, and widespread deployments but did not focus on the performance of libraries.

Different related works analyzed the performance of \quic implementations \cite{yang2020quicnicoffloading,tyunyayev2022picoquichighspeed,yu2021quicproductionperformance,sander2022resourceprioritization,megyesi2016howquickisquic,shreedhar2022quicperformancewebandstorage,wolsing2019performance}.
However, they either analyzed \quic in early draft stages, \eg \citet{megyesi2016howquickisquic} in \citeyear{megyesi2016howquickisquic}, or mainly focused on scenarios covering small web object downloads across multiple streams, \eg by \citet{sander2022resourceprioritization} and \citet{wolsing2019performance}.
Similar to our work, \citet{yang2020quicnicoffloading} orchestrated \quic implementations to download a file.
With a single stream, they reached a throughput between \SI{325}{\Mbps} and even \SI{4121}{\Mbps} for Quant.
However, they omit \ac{http}/3 and only download a file of size \SI{50}{\mega\byte}.
Therefore, the impact of the handshake and cryptographic setup is higher, and other effects might be missed.

\citet{endres2022qirsattelite} adapted the \ac{qir} to evaluate \quic implementations similar to the starting point of our framework.
However, they still relied on the virtualization using docker and network emulation using ns-3 and focused on a different scenario, namely satellite links.

\citet{tyunyayev2022picoquichighspeed} combined \picoquic with the \ac{dpdk} to bypass the kernel.
They compared their implementation to other \quic stacks and increased the throughput by a factor of three.
They argue that the speedup is primarily due to reduced \ac{io} but do not investigate other factors in more detail.

\section{Conclusion}
\label{sec:conclusion}
In this work, we analyzed the performance of different \quic implementations on high-rate links and shed light on various influence factors.
We systematically created a measurement framework based on the idea of the \acl{qir}.
It allows automating \quic goodput measurements between two dedicated servers.
It can use different \quic implementations, automate the server configuration, collect various statistics, \eg from the network device and \ac{cpu} statistics, and provide means to collect, transform, and evaluate results.

We applied the presented framework to evaluate the goodput of mainly \lsquic and \quiche on \SI{10}{\Gbps} links and analyzed what limits the performance.
A key finding in this work is that the \ac{udp} receive buffer is too small by default, which leads to packets getting dropped on the receiver side.
This results in retransmissions and a reduced goodput.
We show that increasing the buffer by at least an order of magnitude is necessary to reduce buffer limits in high link rate scenarios.
We observed several differences in the behavior of \lsquic and \quiche, such as differing default parameters, \eg the \ac{udp} packet size or a diverse approach regarding the acknowledgment sending rate.
When comparing different \ac{tls}~1.3 ciphers, \quic almost reaches the same goodput with ChaCha20 as with the hardware-accelerated \ac{aes} ciphers, which behaves differently with \ac{tcp}.
We could not measure any performance increase with the support of segmentation offloading features of the operating system.
Finally, we show that evaluating \quic highly depends on the used CPU.
By applying various optimizations, we increased the goodput of \lsquic by more than \SI{25}{\percent} %
and achieved up to \SI{5}{\Gbps} on Intel CPUs.

Even though \quic has many similarities to \ac{tcp} and the specification took several years, our work shows that many details already analyzed and optimized for \ac{tcp} are still limiting \quic.
Furthermore, the variety of implementations complicates a universal evaluation and yields further challenges to improve performance in the interplay of libraries, \eg the drastically reduced goodput using an \lsquic server and \quiche client, as shown in \Cref{sec:baseline}.

To allow for an informed and detailed evaluation of \quic implementations in the future, we publish the framework code, the analysis scripts, and the results presented in the paper~\cite{code-publication}.
The measurement framework can be applied to evaluate future improvements in \quic implementations or operating systems.

\section*{Acknowledgment}
\label{sec:acknowledgment}

The European Union's Horizon 2020 research and innovation programme funded this work under grant agreements No 101008468 and
101079774.
Additionally, we received funding by the Bavarian Ministry of Economic Affairs, Regional Development and Energy as part of the project 6G Future Lab Bavaria.
This work is partially funded by Germany Federal Ministry of Education and Research (BMBF) under the projects 6G-life (16KISK001K) and 6G-ANNA (16KISK107).

\pagebreak %

\bibliographystyle{IEEEtranN}
{
\footnotesize
\bibliography{IEEEabrv,paper.bib}
}

\begin{figure}[htb]
	\centering
	\includegraphics{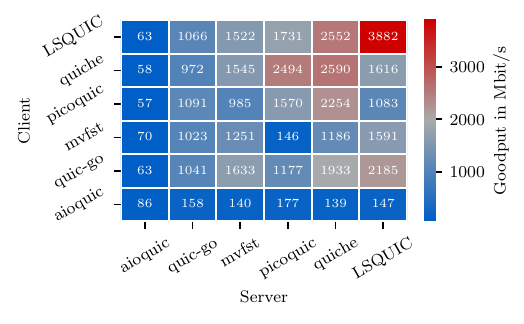}
	\caption{Goodput results for all implementations with increased buffer sizes and optimize compile flags for each implementation.}
	\label{fig:baseline_optimized}
\end{figure}

\appendix
We repeated the baseline measurements presented in \Cref{sec:baseline} but with increased \ac{udp} buffer size and the optimized implementations for \lsquic and \quiche.
As seen in \Cref{fig:baseline_optimized}, \lsquic profits from the optimizations and the larger buffer while other implementations such as \aioquic and \quicgo are limited by other other bottlenecks.

\label{lastpage}

\end{document}